\documentclass[11pt]{article}
\usepackage{amsmath,amssymb,graphicx,epsfig,cite,color,hyperref,booktabs}
\def\smallfrac#1#2{\hbox{${{#1}\over {#2}}$}}
\def\half{\hbox{${1\over 2}$}}
\def\MSb{$\overline{\rm MS}$}

\begin{document}

\vspace{-2.0cm}
\begin{flushright}
Edinburgh 2018/3\\
BNL-200028-2018-JAAM\\
\end{flushright}

\begin{center}
\vspace{2.0cm}

{\Large \bf The Proton Spin, Semi-Inclusive processes, and a future Electron Ion Collider}

\vspace{1.7cm}

Richard~D.~Ball$^{1}$, A.~Deshpande$^{2}$

\vspace{1.3cm}
{\it ~$^1$ The Higgs Centre for Theoretical Physics, University of Edinburgh,\\
JCMB, KB, Mayfield Rd, Edinburgh EH9 3JZ, Scotland\\
~$^2$  Center for Frontiers in Nuclear Science, Stony Brook University,\\ 
Stony Brook, NY 11794-3800\\ 
\& Brookhaven National Laboratory, Upton, NY 11973-5000\\
}

\end{center}   

\vspace{2.1cm}

\begin{center}
  {\bf \large Abstract}
\end{center}

We discuss spin physics, Guido's contribution to it, and what we still have to learn. We set out in particular a programme for incorporating constraints from semi-inclusive data into global fits of polarized PDFs, and discuss the need for the EIC to increase the precision and kinematic coverage of current measurements.

\vspace{2.1cm}
Contribution to the volume {\it ``From my Vast Repertoire: the Legacy of Guido Altarelli"}

\clearpage

\section{Introduction}

Guido Altarelli's interest in spin physics goes right back to the start of the partonic approach to perturbative QCD \cite{Altarelli:1977zs,Altarelli:1978id,Altarelli:1979ub,Altarelli:1979kv}: already in the Altarelli-Parisi paper the structure function $g_1(x,Q^2)$, measurable in inclusive longitudinally polarized deep inelastic scattering, is expressed in terms of polarized parton densities (pPDFs) $\Delta q(x,Q^2)$, $\Delta \bar{q}(x,Q^2)$, $\Delta g(x,Q^2)$, and the corresponding polarized splitting functions are computed in parton language. The results were 
consistent with the LO polarized anomalous dimensions calculated a one year previously in the OPE formalism \cite{Ahmed:1976ee}. Section 4.6 of what became known as Guido's `Bible of QCD' \cite{Altarelli:1981ax} is devoted to the theory of polarized parton densities, and remains recommended reading. 

The EMC experiment~\cite{Ashman:1988}, using a longitudinally
polarized muon beam and a stationary target that contained polarized
protons, was the first experiment to explore $g_1(x,Q^2)$ down to
momentum fractions $x$ as low as $0.01$. When extrapolated over
unmeasured $x < 0.01$ and combined with the couplings in leptonic
hyperon decays and the assumption of $\rm{SU(3)}$ flavor
symmetry, this led to the famous conclusion that the quark and anti-quark spins constitute only a small
fraction of the proton spin.  In addition, with these assumptions, the
polarization of the strange quark sea in the polarized proton was found
to be large and negative.

The resolution of this `spin crisis' by Altarelli and Ross \cite{Altarelli:1988nr} is a beautiful example of a deep theoretical insight of direct relevance to experiment. The nonsinglet first moments of the quark distributions $q^+\equiv q+\bar{q}$, the axial charges $a_3=\Delta u^+ - \Delta d^+$ and $a_8=\Delta u^+ +\Delta d^+ - 2\Delta s^+$, correspond in the OPE to matrix elements of conserved currents, and thus there exist factorization schemes (such as \MSb) in which they are independent of $Q^2$, and can thus be identified with axial decay constants measured in hyperon decays, interpreted through the quark model. The singlet first moment $a_0$ is different however, since the conservation of the singlet axial current is violated by the axial anomaly, and $a_0$ thus evolves in \MSb\ (at two loops). This scale dependence makes the identification of the axial singlet charge $a_0(Q^2)$, extracted from the EMC data, with the total helicity of the quarks problematic, since quark model predictions rely on conserved currents and are thus scale independent. 

However there exist alternative
factorization schemes in which the axial anomaly is not renormalized \cite{Adler:1969er}, and in these schemes  
\begin{equation}
\label{eq:anomaly}
a_0(Q^2) = \Delta\Sigma - n_f\smallfrac{\alpha_s(Q^2)}{2\pi}\Delta g(Q^2), 
\end{equation}
where $\Delta\Sigma$ is the singlet first moment of the polarized quark distributions, and $\Delta g(Q^2)$ is the first moment of the polarized gluon distribution. $\Delta\Sigma$ is independent of $Q^2$, to all orders, and it may thus be identified with $\Delta u^+ +\Delta d^+ +\Delta s^+$, and interpreted as the total helicity of the quarks \cite{Altarelli:1988nr}. Moreover, although in 
Eq.~\ref{eq:anomaly} the second term appears to be of ${\cal{O}}(\alpha_s)$, and thus small, because $a_0(Q^2)$ only evolves at two loops, $\Delta g$ grows as $1/\alpha_s$ at large $Q^2$, and thus the difference between $a_0$ and $\Delta\Sigma$ can be of order one even at large scales. It follows that if the first moment of polarized gluon distribution $\Delta g(Q^2)$ is large and positive, it can compensate $\Delta\Sigma$, reduce $a_0$, and thus explain the EMC measurement.

For a number of years this resolution of the 'spin crisis' was rather controversial, with many people either confirming or contesting the arguments in \cite{Altarelli:1988nr,Altarelli:1988mu,Altarelli:1990jp}, often with more heat than light (see for example \cite{Altarelli:1990wm,Ball:1995de,Forte:1995gs}, and refs therein). This was in part because the issues are rather subtle and the result surprising, but also perhaps because the mathematical structure of chiral anomalies had only recently been unravelled (see ref.\cite{Ball:1988xg} for a contemporary review). A complete two loop calculation \cite{Mertig:1995ny,Vogelsang:1995vh,Vogelsang:1996im} of the \MSb\ polarized splitting functions settled most of the theoretical issues regarding the scheme dependence, and in particular clarified the relation with the Adler-Bardeen condition \cite{Ball:1995td} (polarized splitting functions have recently been computed at three loops \cite{Moch:2014sna,Moch:2015usa}). Some theoretical studies were made of the small $x$ behaviour of the pPDFs \cite{Berera:1992tn,Ball:1995ye,Blumlein:1996sy}, in an attempt to improve the small $x$ extrapolation when computing first moments. However the central questions remained unanswered: is $\Delta g$ large and positive, or does $\Delta s$ violate expectations from $\mathrm{SU(3)}$? 
Since pPDFs are by nature nonperturbative quantities, these questions can only really be answered by experiment.

\section{Current Polarized DIS and p-p Data}

Significant progress has been made since the EMC
observations on the proton's spin
composition.  One main focus has been on measurements with
longitudinally polarized lepton beams scattering off longitudinally
polarized nucleons in stationary targets.  Inclusive data have been
obtained in experiments at
CERN~\cite{Adeva:1998vv,Alexakhin:2006vx,Alekseev:2010hc},
DESY~\cite{Ackerstaff:1997ws,Airapetian:2006vy}, Jefferson
Laboratory~\cite{Zheng:2004ce,Dharmawardane:2006zd}, and
SLAC~\cite{Anthony:1996mw,Abe:1998wq,Abe:1997cx,Anthony:1999rm,Anthony:2000fn}
in scattering off targets with polarized protons and neutrons.  The
kinematic reach and precision of the data on $g_1(x,Q^2)$ so far is
similar to that of the unpolarized structure function $F_2(x,Q^2)$
just prior to the experimental program at the HERA electron-proton
collider. 

Figure~1 provides a survey of the regions in $x$ and $Q^2$ 
covered by the world
polarized-DIS data, which is roughly $0.004 < x < 0.8$ for $Q^2 >
1\,\mathrm{GeV}^2$.  For a representative value of $x \simeq 0.03$,
the $g_1(x,Q^2)$ data are in the range $1\,\mathrm{GeV}^2 < Q^2 <
10\,\mathrm{GeV}^2$.  This is to be compared to $1\,\mathrm{GeV}^2 <
Q^2 < 2000\,\mathrm{GeV}^2$ for the unpolarized data on $F_2(x,Q^2)$
at the same $x$. The figure also shows the vast expansion in $x,Q^2$
reach that an EIC would provide, as will be discussed below.  
Over the past 15 years, an additional powerful line of experimental
study of nucleon spin structure has emerged: {\em semi-inclusive}
deep-inelastic scattering (SIDIS).  In these measurements, a charged or
identified hadron is observed in addition to the scattered
lepton. The relevant spin structure functions then depend on fragmentation functions~\cite{Altarelli:1979kv} which depend on the the momentum fraction that is transferred from the outgoing quark or anti-quark to the observed hadron. The non-perturbative fragmentation functions are at
present determined primarily from precision data on hadron production
in $e^+e^-$ annihilation, but data from the $B$-factories and the LHC are now helping to
further improve their determination~(see~\cite{Albino:2008gy,Metz:2016swz} for recent 
reviews). Insights from the semi-inclusive
measurements are complementary to those from the inclusive
measurements. Specifically, they make it possible to delineate the
quark and anti-quark spin contributions by flavor, since $\Delta q$
and $\Delta \bar{q}$ appear with different weights in the SI cross-section.
A large body of semi-inclusive data sensitive to
nucleon helicity structure has been collected by the experiments at
CERN and
DESY in the last 20 years.

A further milestone in the study of the nucleon was the advent of
RHIC, the world's first polarized proton+proton collider. In the
context of the exploration of nucleon spin structure, the RHIC spin
program is a logical continuation. Very much in the spirit of the
unpolarized hadron colliders in the 1980's, RHIC entered the scene to
provide complementary information on the nucleon that is not readily
available in fixed-target lepton scattering.  The measurement of the
spin-dependent gluon distribution $\Delta g(x,Q^2)$ in the proton is a
major focus and strength of RHIC. Here the main tools are spin
asymmetries in the production of inclusive
pions~\cite{Adler:2004ps,Adare:2007dg,Adare:2008px,Abelev:2009pb,Adare:2014hsq} and
jets~\cite{Abelev:2006uq,Abelev:2007vt, Adamczyk:2014ozi}
at large transverse momentum perpendicular to the beam axis, which
sets the hard scale $Q$ in these reactions. Their reach in $x$ and $Q^2$ is also indicated in
Fig.~1. Unlike DIS, the processes used at RHIC do not
probe the partons locally in $x$, but rather sample over a region in
$x$. RHIC also provides complementary information on
$\Delta u,\Delta \bar{u}, \Delta d,\Delta \bar{d}$ for $0.05 < x <
0.5$~\cite{Adare:2010xa,Aggarwal:2010vc,Adamczyk:2014xyw,Gal:2014fha}, 
with a beautiful technique that exploits the violation of parity and does not rely on knowledge of
fragmentation.  The carriers of the charged-current weak interactions,
the $W$ bosons, naturally select left-handed quarks and right-handed
anti-quarks, and their production in $p$+$p$ collisions at RHIC and
calculable leptonic decay hence provide an elegant probe of nucleon
helicity structure.

\section{Nucleon spin structure and Polarized PDF Fits}

\subsection{Global fits of polarized PDFs}
The earliest attempts \cite{Altarelli:1988mu,Altarelli:1993np,Altarelli:1994ug,Gehrmann:1994rb,Ball:1995ye} to determine polarized PDFs (pPDFs) from the data on inclusive polarized DIS relied on LO perturbation theory, since that was all that was available at the time. Various issues, in particular the $Q^2$ dependence of the asymmetries, and the importance of the small-$x$ extrapolation region, were better understood, and significant improvements in the quality of the data meant that first attempts could be made to determine $\Delta g$ from scaling violations. Following, the publication of the two loop splitting functions 
\cite{Mertig:1995ny,Vogelsang:1995vh,Vogelsang:1996im}, these fits were upgraded to full NLO perturbative QCD
\cite{Ball:1995td,Gehrmann:1995ag,Altarelli:1996nm,Altarelli:1998nb}, and incorporated all the most recent inclusive pDIS data from SMC and E142, E143 and E154 on both proton and deuteron targets. Final results for the first moments were however still inconclusive: the suppression of the axial singlet charge $a_0$ was confirmed, $\mathrm{SU(3)}$ breaking seemed small, and the data showed a mild preference for a positive $\Delta g$, but the uncertainties were too large to allow more definite conclusions.

By this time the difficulties were becoming clear. Reliable PDF determination requires reliable uncertainties, and for this it was necessary to develop new statistical techniques. Parametrization bias was a major issue: this is particularly the case for pPDFs since they can cross over from positive to negative, and thus have quite a complicated shape. On the experimental side, inclusive DIS data is insufficient to determine $\Delta q$ and  $\Delta \bar{q}$ separately, and since strangeness is important, the flavour separation provided by proton and deuteron targets is insufficient. The narrow range in $Q^2$ of the experimental data made extraction of the gluon from scaling violations alone very challenging, while the limited range in $x$ created substantial uncertainty in the extrapolation to small $x$ required to determine first moments. 

These issues have been thrown into sharp relief by the development of more sophisticated techniques for determining global PDFs (see \cite{Ball:2014uwa,Ball:2017nwa,Thorne:2015rch,Dulat:2015mca,Ball:2015oha} and ref therein), with in particular more reliable error estimation through the use of Hessians \cite{Pumplin:2001ct}, Lagrange multipliers \cite{Stump:2001gu}, Monte Carlo replicas \cite{Giele:1998gw} and unbiased parametrizations using neural networks \cite{Forte:2002fg}. Applying the NNPDF methodology, developed for unpolarized PDFs, to the determination of pPDFs from inclusive DIS, showed just how large the uncertainties actually are \cite{Ball:2013lla,Nocera:2015vva}. Even when recent data from COMPASS and JLAB are 
included~\cite{Leader:2014uua,Jimenez-Delgado:2013boa,Sato:2016tuz}, realistically determined uncertainties on $\Delta s^+$ and $\Delta g$ are so large that they are both still consistent with zero. It thus becomes essential to attempt to incorporate other types of data into the determination of pPDFs.

\subsection{Role of SIDIS and Fragmentation Functions}

Semi-inclusive data are useful for separating quark from anti-quark distributions, and giving information on strangeness. Coefficient functions for polarized semi-inclusive DIS (pSIDIS) and single particle hadroproduction \cite{Jager:2002xm} are known at NLO, and there is now a wealth of data from SMC, HERMES and COMPASS, and from PHENIX and STAR. Inclusion of these data in the global fits of pPDFs was first accomplished by DSSV~\cite{deFlorian:2008mr,deFlorian:2009vb} and LSS~\cite{Leader:2010rb}. The DSSV fits seemed to indicate that $\Delta s$ is positive at large $x$, becoming negative at small $x$, and likewise that $\Delta g$ is small, with a first moment consistent with zero. While demonstrating that semi-inclusive data might a significant part to play in the determination of pPDFs, their impact was inevitably limited by the large uncertainties in fragmentation functions (FFs). Indeed, it is for this reason that SI data are rarely included in global determinations of unpolarized PDFs.

The FFs used by DSSV \cite{deFlorian:2007aj,deFlorian:2007ekg} for $\pi^\pm$, $K^\pm$ and $p/\bar{p}$ were determined from semi-inclusive annihilation data (SIA) and unpolarized SIDIS. Although the uncertainties were large, they were not well quantified, and this inevitably led to questions about the reliability of the inclusion of pSIDIS in the determination of pPDFs. In recent years however the situation has improved, with a new generation of FFs \cite{deFlorian:2014xna,deFlorian:2017lwf,Sato:2016wqj,Bertone:2017tyb,Nocera:2017gbk} with more reliable uncertainties. It is thus now possible to employ polarized SI data in a global fit of pPDFs, propagating the uncertainties from the FFs into the pPDFs. A general scheme for combining consistently fits to PDFs, FFs and pPDFs making the best use of available unpolarized and polarized SI data will be set out in the next section.

\subsection{RHIC Spin data and polarized gluon and sea quark PDFs }

\begin{figure}[!t]
\begin{center}
\includegraphics[width=0.32\textwidth]{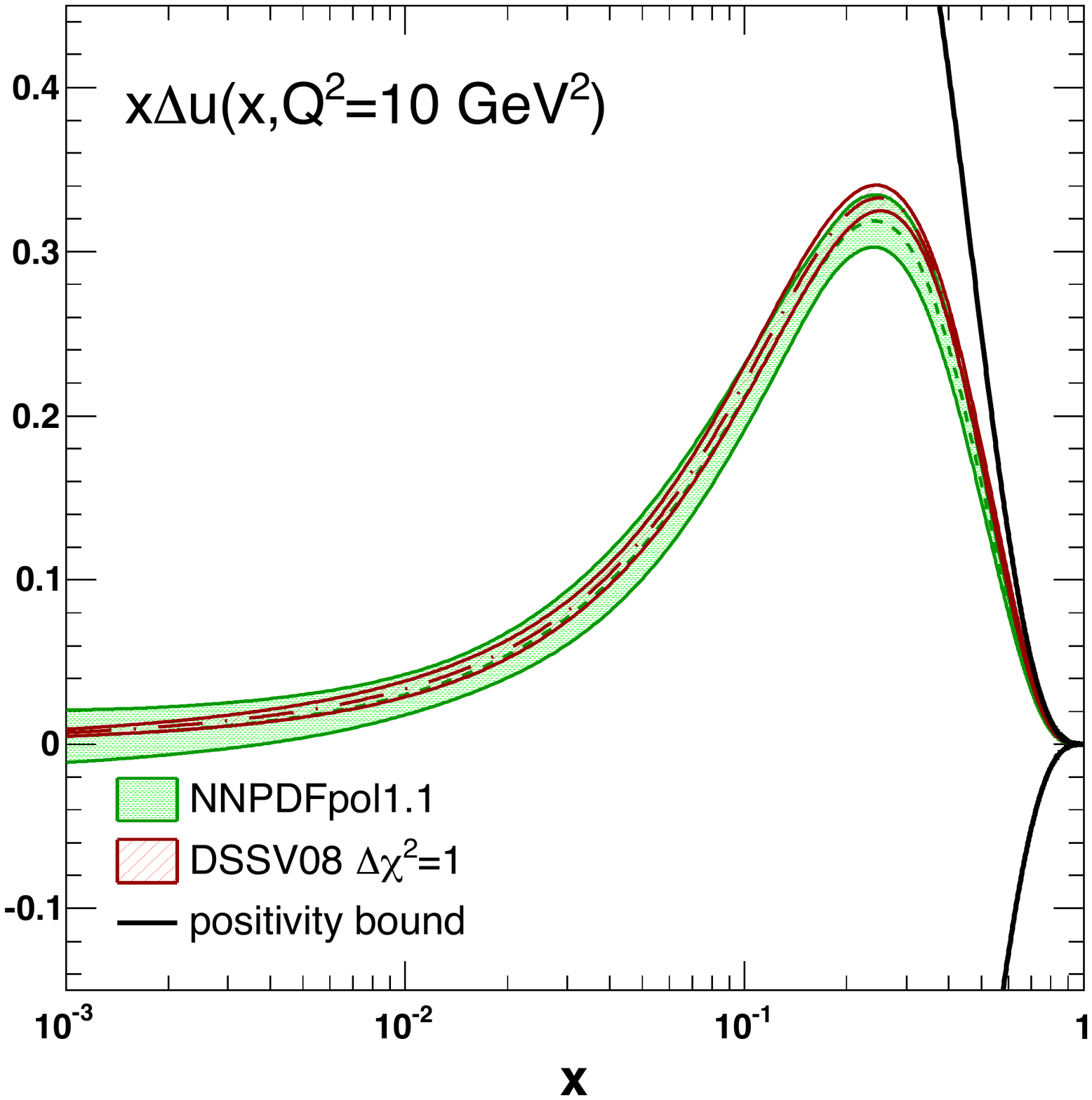}
\includegraphics[width=0.32\textwidth]{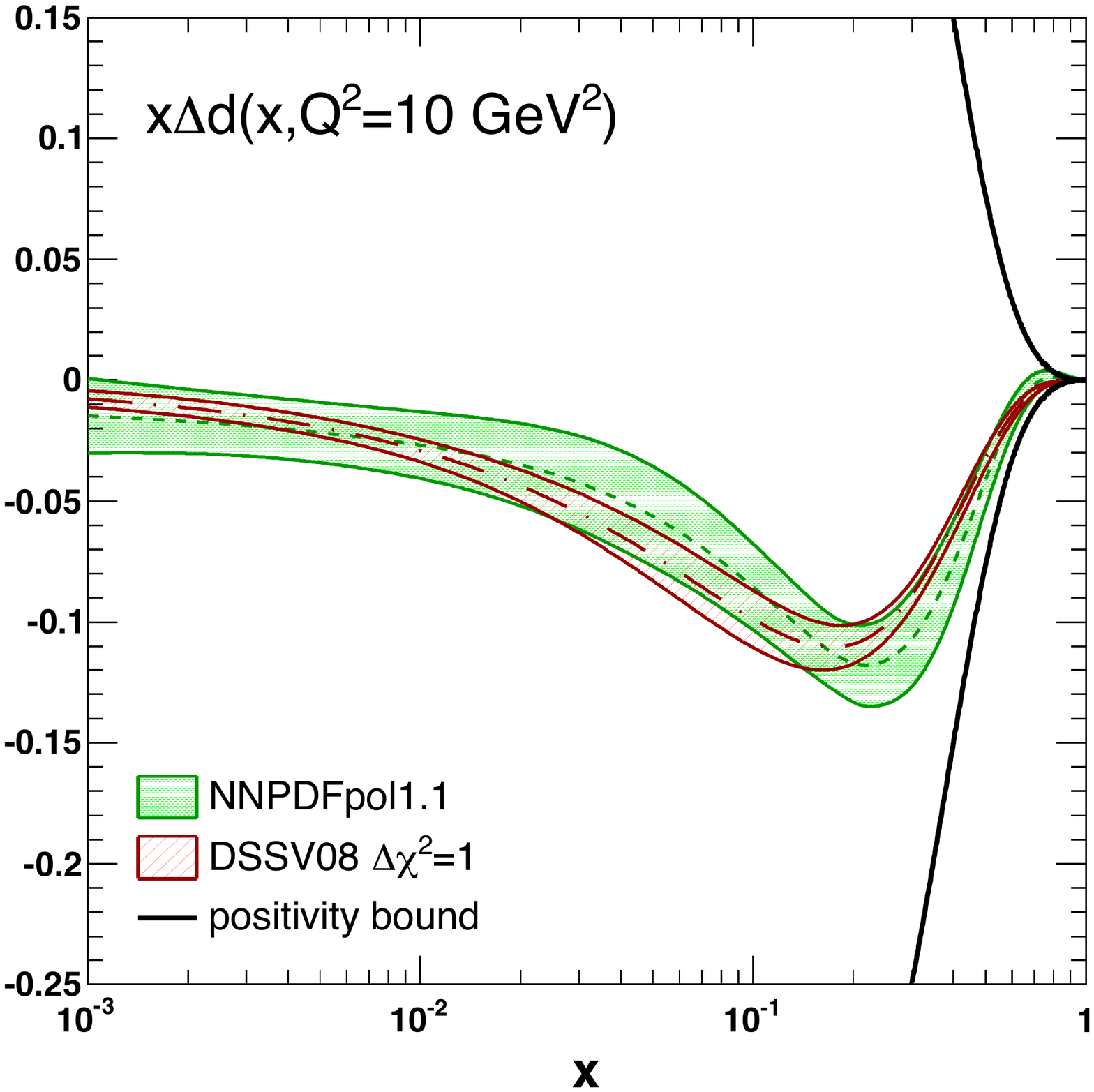}\\
\vskip-2.0cm
\includegraphics[width=0.32\textwidth]{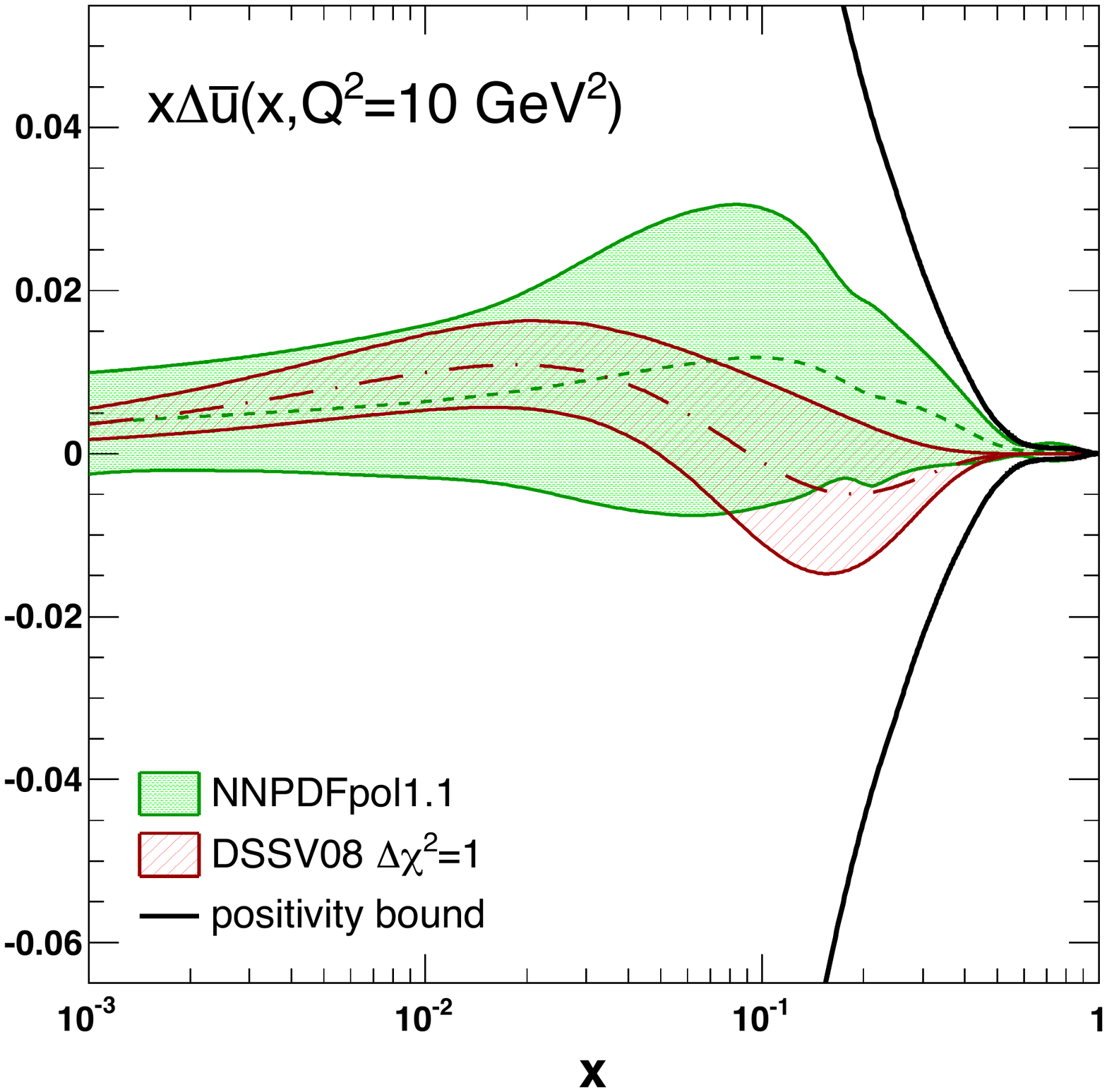}
\includegraphics[width=0.32\textwidth]{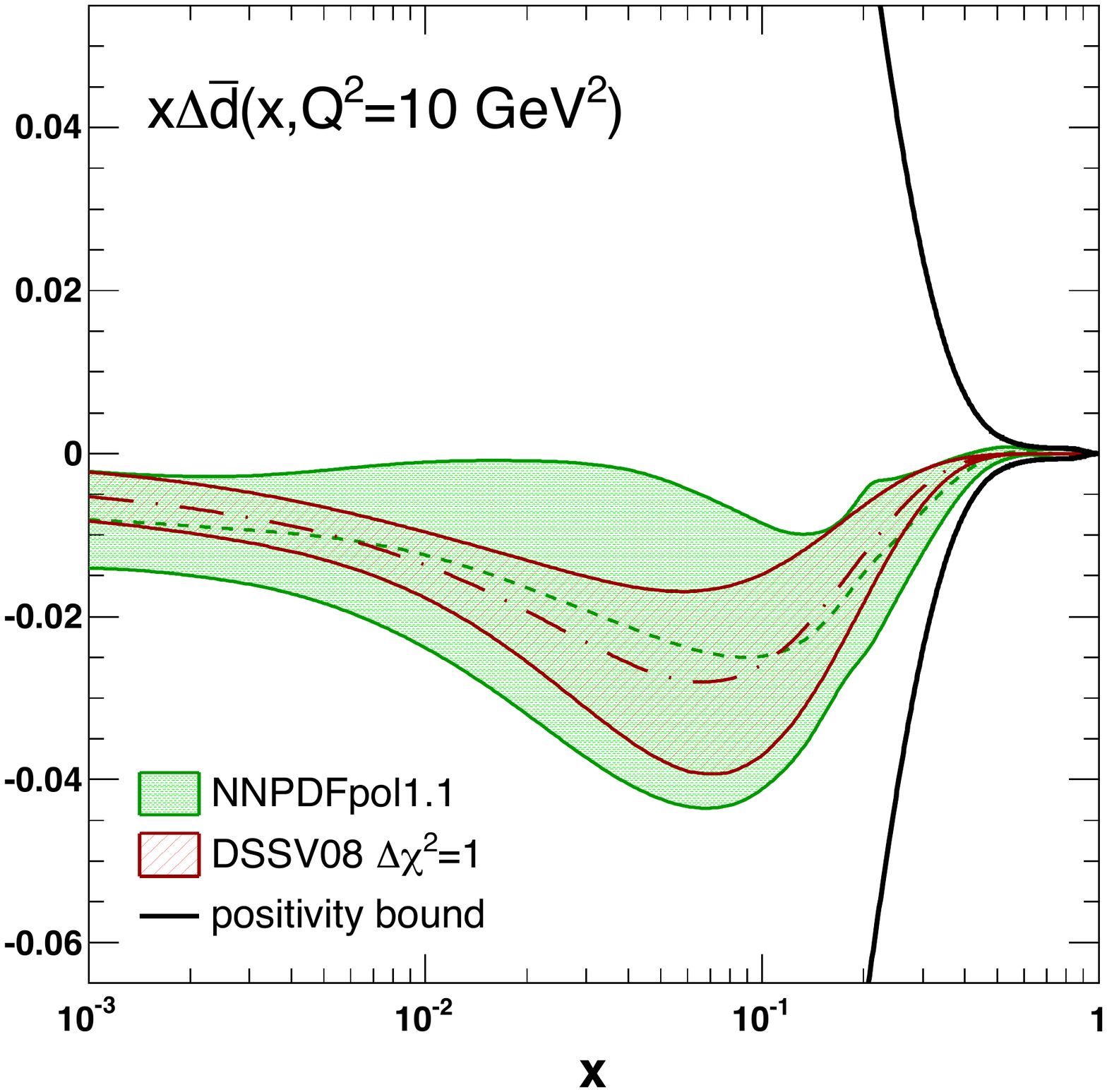}\\
\vskip-2.0cm
\includegraphics[width=0.32\textwidth]{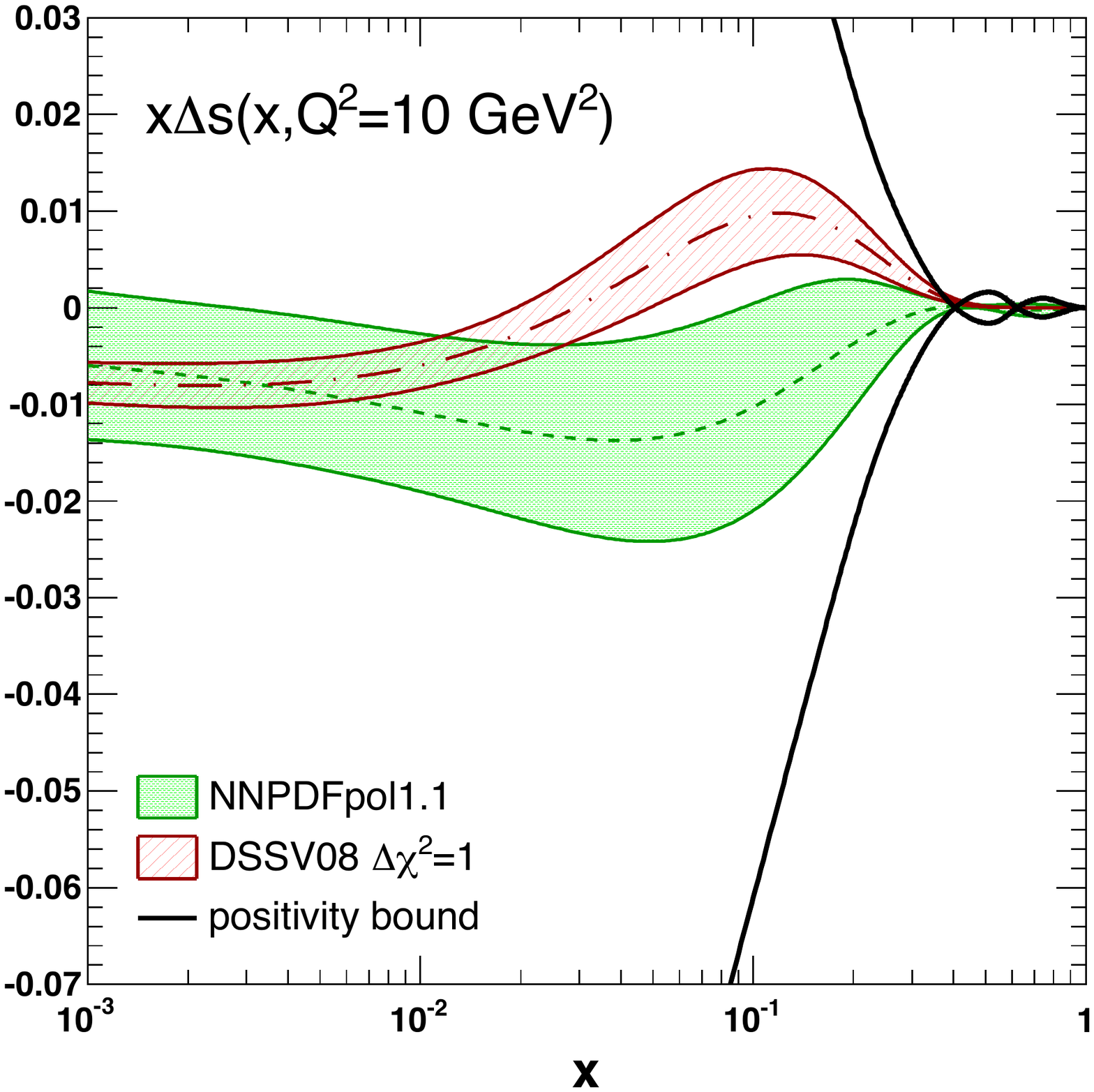}
\includegraphics[width=0.32\textwidth]{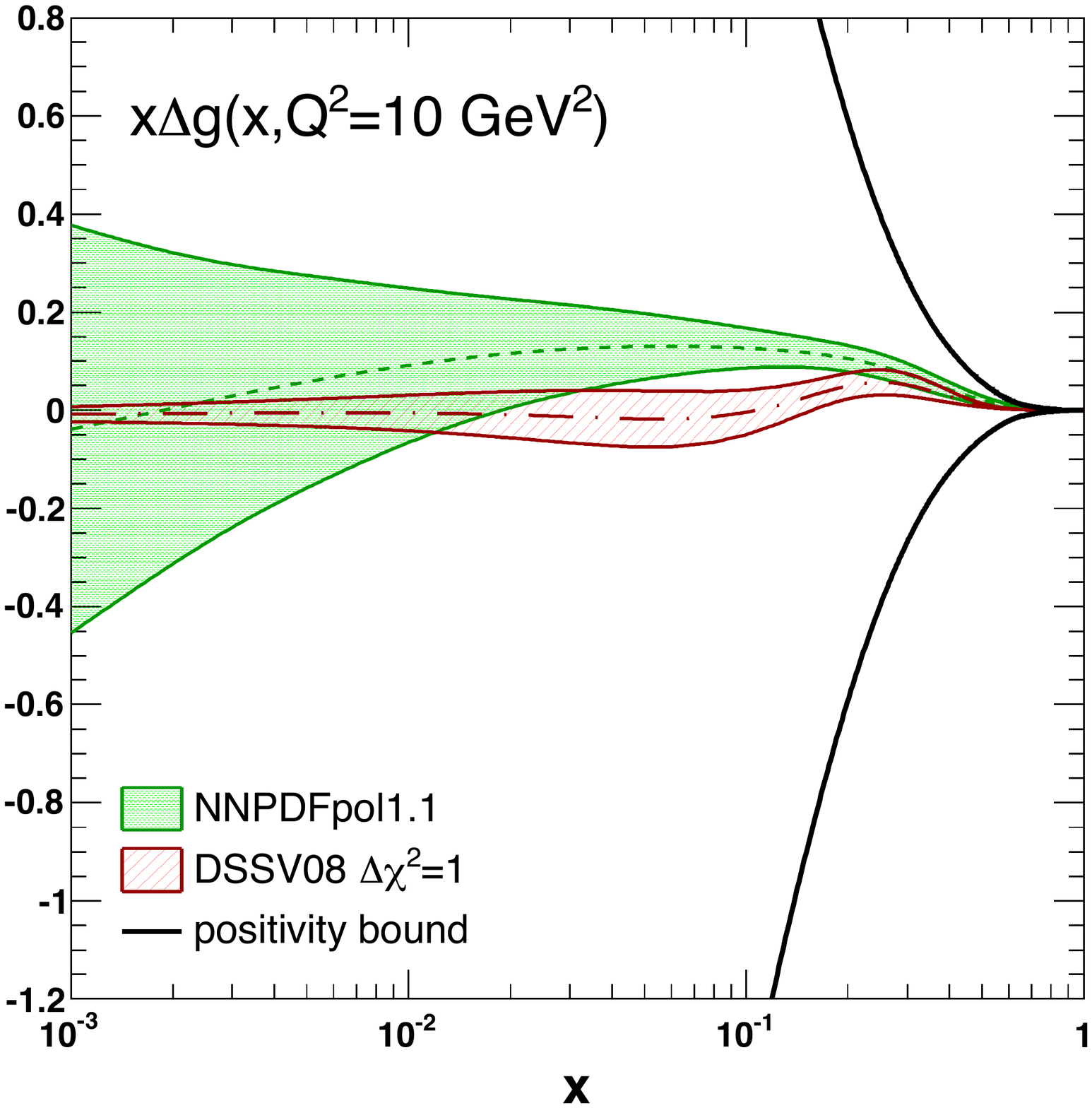}\\
\vskip-1.0cm
\caption{The NNPDFpol1.1 parton set~\cite{Nocera:2017wep} compared to DSSV08~\cite{deFlorian:2009vb} at $Q^{2}=10$ GeV$^{2}$. Both are presented in the \MSb\ scheme.\label{fig:nnpdfs}}

\end{center}
\end{figure}

Perhaps the most significant development in spin physics in recent years has been the advent of Drell-Yan, W/Z production, and inclusive $\pi^{0}$ and jet data from longitudinally polarized proton-proton collisions at RHIC. All these processes can be computed at NLO~\cite{Gehrmann:1997pi,Gehrmann:1997ez,Gluck:2000ek,Jager:2004jh,deFlorian:2010aa}, and as in unpolarized PDF determination they allow us to separate quark and anti-quark distributions, and give a more direct determination of the gluon than is available in inclusive DIS. In recent years truly global determinations of pPDFs have thus become possible \cite{deFlorian:2014yva,Nocera:2014gqa,Nocera:2017wep}, incorporating data on inclusive $\pi^{0}$,  jets and $W^\pm$ production from STAR and PHENIX alongside the inclusive data from pDIS. The pPDFs determined in this way are shown in Fig~\ref{fig:nnpdfs}, 
and some results for first moments are collected in Tab.~\ref{tab:qmom} and Tab.~\ref{tab:gmom}. 
Now at last some evidence for a positive polarized gluon distribution emerges, at least in the measured region. 

\begin{table}[!t]
 \centering
 \small
 \begin{tabular}{c|c|cc}
 \toprule
 & \multicolumn{1}{c|}{$\langle\Delta f\rangle^{[0,1]}$}
 & \multicolumn{2}{c}{$\langle\Delta f\rangle^{[10^{-3},1]}$}\\
 $\Delta f$ 
 & \texttt{NNPDFpol1.1}
 & \texttt{NNPDFpol1.1}
 & \texttt{DSSV08} \\
 \midrule
 $\Delta u^+$
 & $+0.79\pm 0.07$
 & $+0.76\pm 0.04$
 & $+0.793^{+0.028}_{-0.034}\,(+0.020)$\\
 $\Delta d^+$
 & $-0.47\pm 0.07$
 & $-0.41\pm 0.04$
 & $-0.416^{+0.035}_{-0.025}\,(- 0.042)$\\
 $\Delta\bar{u}$
 & $+0.06\pm 0.06$ 
 & $+0.04\pm 0.05$
 & $+0.028^{+0.059}_{-0.059}\,(+ 0.008)$\\
 $\Delta\bar{d}$
 & $-0.11\pm 0.06$
 & $-0.09\pm 0.05$
 & $-0.089^{+0.090}_{-0.080}\,(- 0.026)$\\
 $\Delta s$
 & $-0.07\pm 0.05$
 & $-0.05\pm 0.04$
 & $-0.006^{+0.028}_{-0.031}\,(- 0.051)$\\
 $a_0$
 & $+0.18\pm 0.21$
 & $+0.25\pm 0.10$
 & $+0.366^{+0.042}_{-0.062}\,(+ 0.124)$\\
 \bottomrule
 \end{tabular}
\caption{Full and truncated first moments of the polarized quark distributions at $Q^2=10$ GeV$^2$, for \texttt{NNPDFpol1.1}~\cite{Nocera:2017wep}, \texttt{NNPDFpol1.0}~\cite{Nocera:2014gqa}(based on inclusive DIS data only) and \texttt{DSSV08}~\cite{deFlorian:2009vb} (which included SIDIS data). The uncertainties shown are one-sigma for  NNPDF and  Lagrange multiplier with $\Delta\chi^2/\chi^2=2\%$ for DSSV. The  number in parenthesis for \texttt{DSSV08} is the contribution  that should be added to the truncated moment in order to obtain the full moment. All results are in \MSb\ scheme. \label{tab:qmom}}

\end{table}

\begin{table}[!ht]
 \centering
 \small
 \begin{tabular}{l|ccc}
 \toprule
 & $\langle \Delta g\rangle^{[0,1]}$
 & $\langle \Delta g\rangle^{[10^{-3},1]}$
 & $\langle \Delta g\rangle^{[0.05,0.2]}$ \\
 \midrule
 \texttt{NNPDFpol1.1}
 & $+0.03\pm 3.24$
 & $+0.49\pm 0.75$
 & $+0.17\pm 0.06$\\ 
 \texttt{DSSV08}
 & ---
 & $0.01^{+0.70}_{-0.31}\,(+ 0.10)$
 & $0.01^{+0.13}_{-0.16}$\\
 \texttt{DSSV++}
 & ---
 & ---
 & $0.10^{+0.06}_{-0.07}$\\ 
 \bottomrule
 \end{tabular}
\caption{Same as Tab.~\ref{tab:qmom} but for the \MSb\ gluon. Results are also shown for the truncated moment in the range of the RHIC jet data, for \texttt{NNPDFpol1.1}~\cite{Nocera:2017wep}, \texttt{DSSV08}~\cite{deFlorian:2009vb}, and \texttt{DSSV++}~\cite{deFlorian:2014yva}. \label{tab:gmom}}
\end{table}

The combination of the large body of inclusive deep-inelastic
scattering data off targets containing polarized protons and neutrons
has established that the up quarks and anti-quarks combine to have net
polarization along the proton spin, whereas the down quarks and
anti-quarks combine to carry negative polarization.  The 
$\Delta u^+$ and $\Delta d^+$ distributions are very well constrained by now at medium to large $x$ (see Fig~\ref{fig:nnpdfs}).

The light sea quark and anti-quark distributions still carry
large uncertainties, even though there are some constraints from 
semi-inclusive data and, most recently,  from measurements of spin-dependence 
in leptonic $W$ decay in $\sqrt{s} = 500\,\mathrm{GeV}$ polarized proton+proton
collisions at RHIC~\cite{Adamczyk:2014xyw,Gal:2014fha}. RHIC probes the 
$\Delta u$, $\Delta d$, $\Delta \bar{u}$ and $\Delta \bar{d}$ densities
for $0.05 < x < 0.5$ at a scale set by the $W$-mass.  
The sea shows hints of not being SU(2)-flavor symmetric: the $\Delta \bar{u}$ 
distribution has a tendency to be mainly positive, while the $\Delta \bar{d}$ 
anti-quarks carry opposite polarization, see
Tab~\ref{tab:qmom}. This pattern has been predicted at least 
qualitatively by a number of models of the nucleon. Better constraints on $\Delta u$,
$\Delta d$, $\Delta \bar{u}$ and $\Delta \bar{d}$ are 
are anticipated~\cite{deFlorian:2010aa} from additional RHIC measurements with 
higher integrated luminosity. Furthermore the large luminosities and high
resolution available at the Jefferson Laboratory after an upgrade to
$12\,\mathrm{GeV}$ electron beam energy will extend the kinematic
reach of the existing JLab inclusive and semi-inclusive
deep-inelastic scattering data to twice smaller $x$ as well as to
larger $x$ than have thus far been measured.

Strange quarks appear to be deeply involved in nucleon spin
structure. As we mentioned earlier, from the inclusive deep-inelastic
data, along with SU(3) flavor symmetry considerations, one finds a
negative value for the integrated strange helicity distribution.
Strange quarks and anti-quarks would thus be polarized opposite to the
nucleon.  This would need to be viewed as part of the reason why the
total quark and anti-quark spin contribution $\Delta\Sigma$ is so much smaller
than expected in simple models. A variety of mechanisms for $\mathrm{SU(3)}$
flavor symmetry breaking have been discussed over the years~\cite{Savage:1996zd,Zhu:2002tn,Cabibbo:2003cu,Ratcliffe:2004jt,Sasaki:2008ha}. 
The semi-inclusive measurements with identified
kaons~\cite{Alekseev:2010ub,Airapetian:2004zf} are hence of particular
interest since they potentially yield the most direct measurements of 
the polarization of strange quarks and anti-quarks, albeit with
considerable dependence on the kaon fragmentation
functions~\cite{Leader:2011tm}.  No evidence for sizable $\Delta
s(x,Q^2)$ or $\Delta \bar{s}(x,Q^2)$ has so far been found in polarized
semi-inclusive measurements with fixed targets (see
Tab~\ref{tab:qmom}). As a consequence,
$\Delta s$ would need to obtain its negative integral purely from the
contribution from the thus far unmeasured small-$x$ region. This situation 
demonstrates rather clearly the
need for simultaneous measurements of the kaon production
cross-sections and their spin-dependence in semi-inclusive
deep-inelastic scattering at smaller $x$.

Constraints on the spin-dependent gluon distribution $\Delta g$
predominantly come from RHIC, with some information also entering from
scaling violations of the deep-inelastic structure function
$g_1(x,Q^2)$.  The production cross sections for inclusive hadrons and
jets at RHIC receive contributions from gluon-gluon and quark-gluon
scattering and probe $\Delta g(x,Q^2)$ over the range $0.02 < x <
0.4$. Note that the $x$ is not explicitly resolved
in measurements of inclusive pion and jet probes. Initial results
from RHIC saw small double-spin asymmetries for inclusive jets
and hadrons. As a result, the older DSSV fits~\cite{deFlorian:2008mr,deFlorian:2009vb} 
concluded that there were no indications of a sizable contribution of gluon spins to the
proton spin. The latest much more precise RHIC data
for the double-spin asymmetry in $\pi^{0}$ and jet production~\cite{Adamczyk:2014ozi}
provide, for the first time, evidence of a non-vanishing polarization of
gluons in the nucleon in the RHIC kinematic regime~\cite{deFlorian:2014yva,Nocera:2017wep}, see
Tab~\ref{tab:gmom}.

The limited $x$-range and unresolved small-$x$ dependence still preclude definitive
conclusions on the total gluon spin contribution to the proton spin,
$\Delta_g$, although it appears likely now that gluons are an important player. 
Final results from RHIC at $\sqrt{s} = 200\,\mathrm{GeV}$ will
enhance the sensitivity primarily at large $x$, and measurements of
correlated probes are anticipated to yield insights in $x$-dependence.
Forthcoming measurements at $\sqrt{s} = 500\,\mathrm{GeV}$ are
expected to extend the small-$x$ reach to $2\div3$ times smaller
values and modest further gains may be possible with new instruments
at larger pseudorapidity.  

\begin{figure}[ht]
\label{fig:xq2}
\centering
\includegraphics[width=80mm]{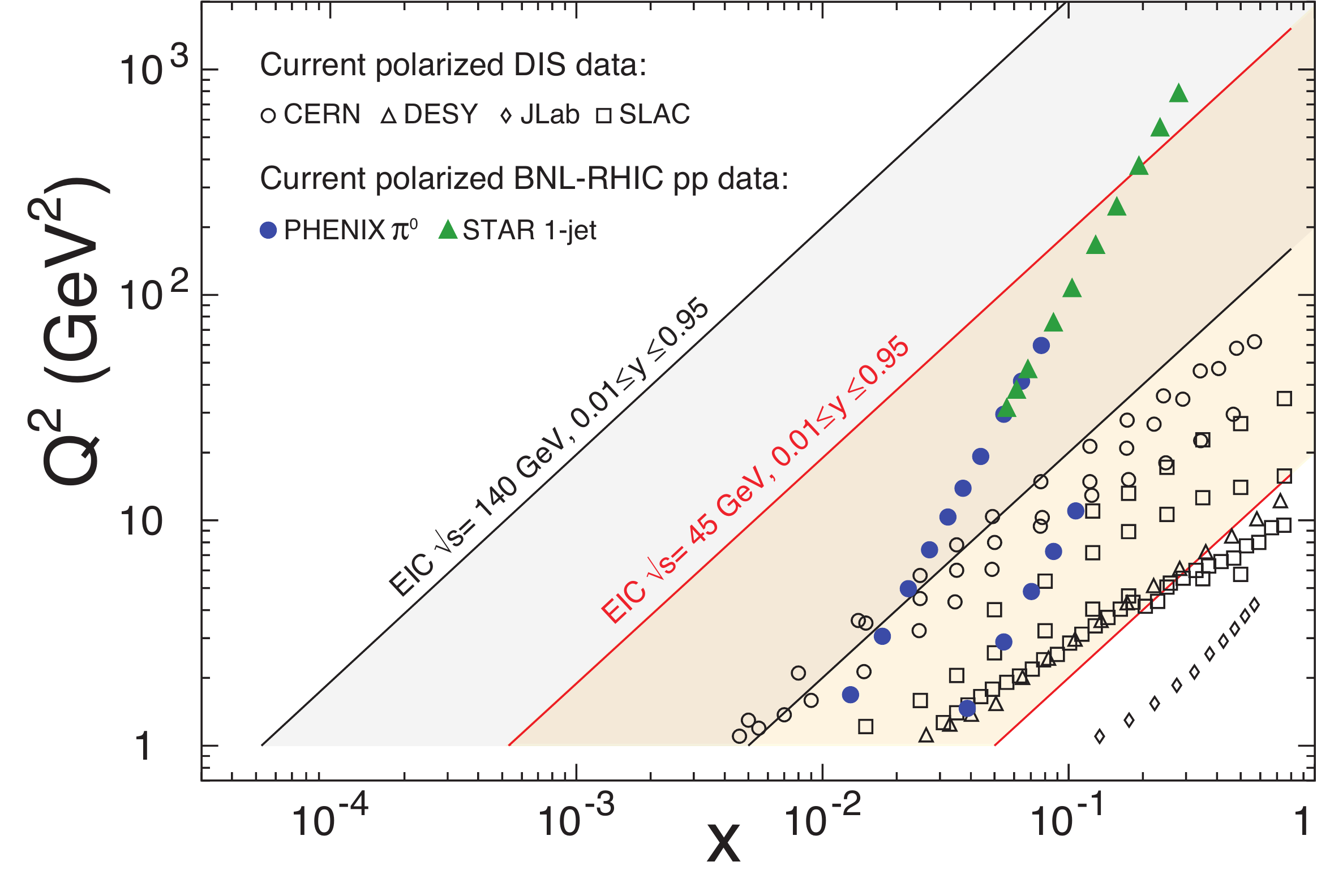}
\caption{Kinematic reach of the current data from fixed target electron and muon scattering experiments around the world (SLAC, JLab and CERN). Also shown for comparison are from RHIC at a scale $Q^2 = p_{T}^{2}$ where $p_{T}$ is the observed jet (pion) transevrse momentum and an $x$ value that is representative for the measurement scale. The $x$ ranges probed at different scales are wide and have considerable overlap. Area covering the $x-Q^2$ reach of the future Electron Ion Collider are also shown for the minimum and maximum $\sqrt{s}$ of $45$ and $140$ GeV.}
\end{figure}

While the data from RHIC have had a tremendous impact, particularly as they improve in precision, the essential difficulty in determining the first moments $\Delta\Sigma$ and $\Delta g$ in Eq.~(\ref{eq:anomaly}) remains: the kinematic reach of current experiments is still insufficient to reliably extrapolate from the measured region to small $x$. We now have tantalizing indications that both $\Delta s$ and $\Delta g$ may become negative at small $x$, with obvious implications for their first moments, so this is a key issue. Theoretically, the perturbative expansion of the polarised splitting functions becomes increasingly unstable at small $x$~\cite{Ball:1995ye}, with double logarithms of the form $\smallfrac{\alpha_s^n}{x}\log^{2n-2}\smallfrac{1}{x}$ for $n=1,2,\ldots$ (see \cite{Ermolaev:2009cq} and ref therein). Although leading log resummations exist, it is unlikely that they could increase the precision of the pPDFs extrapolated to small-$x$ sufficiently to be able to determine a reliable first moment. Thus all our hopes rest on improved measurements, at a new machine such as the EIC. We discuss this possibility, and what it might achieve, in the last section of this article.

\section{Determining polarized PDFs from SI Processes}

In this section we will consider the incorporation of SI processes in determinations of parton distribution functions (PDFs) and polarized parton distribution functions (pPDFs), through the consistent determination of fragmentation functions (FFs). Guido was always rather sceptical of the use of SI data in PDF fits, for the simple reason that he remained unconvinced that the uncertainties due to the FFs were properly accounted for. Here we will attempt to convince him, belatedly perhaps, that we now have the tools to do this properly. In particular we will show how uncertainties can be propagated consistently through the fits, in such a way that the final results are all mutually self-consistent. The discussion will be described in the NNPDF framework, using a replica representation \cite{Giele:1998gw} of PDFs, pPDFs and FFs, but is easily adapted to Hessian representations. We begin with a general discussion of theoretical uncertainties in PDF fits, and then use the results to discuss SI processes. A general fitting strategy is outlined at the end. 

\subsection{Theoretical Uncertainties in PDF fits}
\label{subsec:theory}

We denote the $n$ data in a particular process by a vector $y$, and the corresponding theoretical predictions by a vector $t[f]$, where $f$ is a generic ensemble of PDF replicas. Then by Bayes Theorem
\begin{equation}
  P(t|yf)P(y|f) = P(y|tf)P(t|f),
 \label{eq:bayes}
\end{equation}
so that if we integrate over all possible predictions,
\begin{equation}
  P(y|f) = \int d^n t\, P(y|tf)P(t|f).
  \label{eq:margin}
\end{equation}
Now if we assume Gaussian uncertainties for the data, $P(y|tf)\sim \exp(-\half\chi^2)$, where the tilde means `up to a normalization factor', and  
\begin{equation}
  \chi^2(y,t,\sigma) = (y-t)^T \sigma^{-1}(y-t),
\label{eq:chi2I}
\end{equation}
is the usual $\chi^2$ function for the inclusive data, $T$ denoting the transpose vector, and summation over the $n$ data points being implicit. The $n$ by $n$ matrix $\sigma$ is the complete covariance matrix for the data $y$, including both statistical and systematic uncertainties. Likewise we also assume the theoretical uncertainties are Gaussian,
\begin{equation}
  P(t|f) \sim \exp( -\half(t-t[f])^T s^{-1}(t-t[f])),
\label{eq:theory}
\end{equation}
where again summation over the data points is implicit, and $s$ is the covariance matrix for the theoretical uncertainties. Substituting these two Gaussians into Eq.~(\ref{eq:margin}), the integrand is  Gaussian, and we find after some algebra that again $P(y|f)\sim \exp(-\half\chi^2[f])$, but now with
\begin{equation}
  \chi^2[f] = \chi^2(y,t[f],\sigma+s) = (y-t[f])^T (\sigma+s)^{-1}(y-t[f]),
\label{eq:chi2It}
\end{equation}
Thus to include Gaussian theoretical uncertainties in a PDF fit, all we need to do is take our estimate $s$ for the covariance matrix of the theoretical uncertainties in the space of the data, and add it to the experimental covariance matrix $\sigma$ when computing the $\chi^2$ function to be maximised in the fit. This result is with hindsight rather obvious: there is no distinction mathematically between an experimental and a theoretical uncertainty, so both should be included together on the same footing. In the limit when theoretical uncertainties are ignored, $s\to 0$, Eq.~(\ref{eq:theory}) becomes a $\delta$-function, and Eq.~(\ref{eq:chi2It}) reduces immediately to the usual expression Eq.~(\ref{eq:chi2I}) in which the only uncertainties are experimental.

It remains to estimate the theoretical covariance matrix $s$. A number of sources of theoretical uncertainty might be envisaged as commonly arising in a PDF fit:\\
(i) statistical theoretical uncertainties from Monte Carlo generators. These are perhaps the easiest to treat, since the matrix $s$ is diagonal, and the uncertainties can either be read off from the MC generator itself, or else estimated by studying the point-to-point fluctuations of the predictions\cite{Ball:2017nwa,Harland-Lang:2017ytb} \\
(ii) systematic theoretical uncertainties from, for example, missing higher order perturbative corrections. Here the covariance matrix might be estimated by varying renormalization and/or factorization scales: then
\begin{equation}
  s^{\textrm{mhou}}_{ij} = \langle (t[f;\mu] - t[f;\mu_0])_i(t[f;\mu] - t[f;\mu_0])_j\rangle,
\label{eq:scale}
\end{equation} 
where $\mu_0$ is the scale adopted in the central prediction (for example $Q$ in DIS), and the angled brackets denote averaging over a given range of $\mu$ according to some prescription. Of course this estimate is subject to all the usual caveats of scale variation, and in particular will depend explicitly on the range of scale adopted, and assumptions regarding the independence (or otherwise) of scales in different processes.\\
(iii) systematic theoretical uncertainties from, for example nuclear corrections. Here one might adopt a procedure similar to (ii), but now averaging over different models of the corrections. Again the result will be only qualitative, but should at least capture the correlations of the corrections between different values of $x$ and $Q$. Alternatively one might use fits to nuclear data, particular when these are quite well determined (for example using data from LHC collisions of heavy nuclei).\\
(iv) in determinations of polarized PDFs, there is a further source of theoretical uncertainty, namely that due from the extraction of the polarized cross-sections from the asymmetries, which requires knowledge of the corresponding unpolarized cross-section. Here the covariance matrix, to be added to the experimental covariance matrix for the extracted polarized cross-sections, could be computed by averaging an expression similar to Eq.~(\ref{eq:scale}) over replicas of the unpolarized PDFs. Uncertainties in positivity constraints\cite{Altarelli:1998gn} could be handled in the same way.

In the next section, we will consider a fifth example, the systematic theoretical uncertainties in semi-inclusive processes from fragmentation functions. 

\subsection{Uncertainties in PDFs and pPDFs due to uncertainties in FFs}
\label{subsec:FFs}

Now consider semi-inclusive processes, in which the theoretical prediction $t[f,d]$ depends  on both PDFs 
$f$ and FFs $d$. One way to proceed would be to attempt to determine $f$ and $d$ simultaneously through a minimization of  $\chi^2[f]+\tilde{\chi}^2[f,d]$, where
\begin{equation}
  \tilde{\chi}^2[f,d]\equiv \chi^2(\tilde{y},\tilde{t}[f,d],\tilde{\sigma}),
\label{eq:chi2fd}
\end{equation}
with the $\chi^2$ function given by Eq.~(\ref{eq:chi2I}), and we denote all SI quantities by adding a tilde. For simplicity from now on the usual sources of theoretical uncertainty will be absorbed into the experimental uncertainties, so that we can focus on the FFs.

There are a number of difficulties with this approach:\\
{\bf (a) Theoretical:} in perturbative QCD, both PDFs and FFs are factorized from the hard cross-section, and are thus universal, in the sense that they are independent of any particular process \cite{Altarelli:1978id}.  This suggests that they must be mutually independent, and indeed this is the way they are always used. (Note that generalised crossing, while being useful to relate perturbative cross-sections, is not expected to hold nonperturbatively, and thus cannot be used to relate PDFs and FFs~\cite{Altarelli:1979kv}.) Determining PDFs and FFs simultaneously by minimising Eq.\ref{eq:chi2fd} would however induce nonuniversal correlations between them. Simply ignoring these correlations would affect the reliability of the residual PDF and FF uncertainties.\\
{\bf (b) Statistical:} both PDFs and FFs enter the cross-section multiplicatively. This means that an uncertainty in the FFs induces a multiplicative uncertainty in the PDFs (and vice versa). Accordingly, the value of the FFs rescales uncertainties in the PDFs, meaning that lower values of the FFs will be preferred. This is a rather subtle version of the d'Agostini bias\cite{DAgostini:1993arp}, and seems unavoidable in a joint minimization.\\
{\bf (c) Computational:} besides the technical complications of a joint minimization, evaluation of $\tilde{t}[f,d]$ is computationally intensive, because of the extra convolutions: while inclusive DIS or SIA have only one convolution per iteration per data point, SIDIS has two (the same as an inclusive hadronic process), while a SI hadronic process has three. This would probably prohibit the use of SI hadronic processes in a fit using lookup tables, as the tables would be very large, and the fitting correspondingly slow. 

The solution to all of these problems is to fit PDFs and FFs independently, but consistently, by treating the uncertainties in one or the other as theoretical uncertainties in the fit. Consider for example adding SI processes to a global PDF fit using only inclusive processes, with FFs already determined using SIA. Then 
for a given fixed PDF the theoretical uncertainty on $\tilde{t}[f,d]$ due to the FFs is 
\begin{equation}
  \tilde{s}^d_{ij}[f] = \smallfrac{1}{N}\sum_d (\tilde{t}[f,d]) - \tilde{t}[f,d_0])_i(\tilde{t}[f,d]) - \tilde{t}[f,d_0])_j,
\label{eq:covd}
\end{equation} 
where the sum is over the $N$ FF replicas, and $d_0$ is the central value of the FF, i.e. replica zero (or, in a Hessian approach, a similar expression summing over Hessian eigenvectors). We then minimise $\chi^2[f]+\tilde{\chi}^2[f,d_0]$ to determine PDF replicas $f$, where the inclusive $\chi^2$ is given by Eq.~\ref{eq:chi2I} (or Eq.~\ref{eq:chi2It} if theoretical uncertainties are also included), and the semi-inclusive contribution
\begin{equation}
 \tilde{\chi}^2[f,d_0]= \tilde{\chi}^2(\tilde{y},\tilde{t}[f,d_0],\tilde{\sigma}+\tilde{s}^d[f_0]), 
\label{eq:chi2fd0}
\end{equation}
the FF uncertainty being treated as a theoretical uncertainty. Note that employing $\tilde{s}^d[f]$ in $\tilde{\chi}^2$ would lead to a d'Agostini bias, so instead we use $\tilde{s}^d[f_0]$, just as in the unbiased treatment\cite{Ball:2009qv} of multiplicative experimental uncertainties, where $\sigma[f]$ is replaced by $\sigma[f_0]$. The value of $f_0$ can then be iterated to self consistency in the usual way. 

Now this procedure is evidently symmetrical: we can use the same SI data to instead improve the FFs, by computing the PDF uncertainties 
\begin{equation}
  \tilde{s}^f_{ij}[d] = \smallfrac{1}{N}\sum_f (\tilde{t}[f,d]) - \tilde{t}[f_0,d])_i(\tilde{t}[f,d]) - \tilde{t}[f_0,d])_j,
\label{eq:covf}
\end{equation} 
and determining the FFs through replicas determined by minimising $\chi^2[d]+\tilde{\chi}^2[f_0,d]$, where now $\chi^2[d]$ is the $\chi^2$ for SIA, given by a formula analogous to Eq.~\ref{eq:chi2I}, and the SI data contribute
\begin{equation}
 \tilde{\chi}^2[f_0,d]= \chi^2(\tilde{y}\tilde,\tilde{t}[f_0,d],\tilde{\sigma}+\tilde{s}^f[d_0]),
\label{eq:chi2f0d}
\end{equation}
with the PDF uncertainties now being treated as theoretical. Indeed it probably makes more sense to start this way around, since as noted previously SIA is not sufficient to determine the charge separation of the FFs, while the PDFs are already well known from the global fit to inclusive data. 

Having obtained a first estimate of FFs in this way, we can of course iterate back and forth, until both FFs and PDFs are determined self consistently: starting from $(f^{(0)},d^{(0)})$ determined independently from the global inclusive PDF and SIA fits respectively, we can use the SI data to improve the FFs, then use these FF to update the PDFs, then the updated PDFs to improve the FFs: $(f^{(0)},d^{(0)})\to (f^{(0)},d^{(1)})
\to (f^{(1)},d^{(1)})\to\cdots$, updating the covariance matrices as we go. It is clear that since after the intial step the further effect of the SI data will be small, this process will converge very rapidly. Moreover the resulting PDF and FF replicas will be statistically independent, as required for universality, there is no d'Agostini bias, and since at each stage only one or the other is being fitted, the convolution with the unfitted distributions can be precomputed before each fit. Consequently the joint determination should take little longer than conventional PDF and FF fits, and most importantly need no new technology.

Finally, once we have determined the FFs in this way, we can use them to improve the global determination of the polarized PDFs, by again minimising $\chi^2[\Delta f]+\tilde{\chi}^2[\Delta f,d_0]$, but now with 
\begin{equation}
\chi^2[\Delta f] = \chi^2(y,t[\Delta f],\sigma) 
\label{eq:chi2Ipt}
\end{equation}
the $\chi^2$ for the inclusive polarized data, $t[\Delta f])$ being the usual polarized predictions for inclusive processes, and 
\begin{equation}
 \tilde{\chi}^2[\Delta f,d_0]= \chi^2(\tilde{y},\tilde{t}[\Delta f,d_0],\tilde{\sigma}+\tilde{s}^d[f_0]), 
\label{eq:chi2Delfd}
\end{equation}
the contribution from the polarized SI data. Again the procedure might be iterated, now between polarised PDFs and fragmentation functions using 
\begin{equation}
\tilde{\chi}^2[\Delta f_0,d]= \chi^2(\tilde{y},\tilde{t}[\Delta f_0,d],\tilde{\sigma}+\tilde{s}^{\Delta f}[d_0]), 
\label{eq:chi2dDelf}
\end{equation}
but this seems unlikely to lead to significant further improvement in the FFs when one remembers that polarized SI data are rather harder to obtain than unpolarized, and thus inevitably have larger uncertainties.  

Note that when determining pPDFs from SI data, there is a significant advantage in fitting the asymmetries rather than extracted polarized cross-sections or structure functions: in the asymmetries not only do many experimental systematics cancel, but also much of the dependence on the FFs will also cancel. The contribution to the covariance matrix from FF uncertainties (the polarized equivalent of Eq.~(\ref{eq:covd})) will thus be significantly smaller for the SI asymmetries than for the polarized cross-sections. Of course it will also be necessary to add a contribution to the experimental covariance matrix from the uncertainties in the unpolarized PDFs, using Eq.~(\ref{eq:covf}), but this should be smaller still.

\subsection{A Strategy for the use of SI data to determine pPDFs}

Putting all this together, we might use the following strategy to improve the determination of pPDFs using SI data. Ideally this should be done using a common methodological framework for determining PDFs, FFs and pPDFs, to ensure consistency of the uncertainties passed between them. Fortunately such a framework now exists: we can use as a baseline global PDFs (determined from a wide range of inclusive DIS and hadronic data) from NNPDF3.1\cite{Ball:2017nwa}, FFs (determined from SIA data) from NNFF1.0\cite{Bertone:2017tyb}, and global pPDFs (determined from polarized inclusive DIS and hadronic data) from NNPDFpol1.1\cite{Nocera:2014gqa}, all determined using NLO pQCD. Alternatively, we might start with the DSSV FFs \cite{deFlorian:2014xna,deFlorian:2017lwf}, and pPDFs~\cite{deFlorian:2014yva}. We then proceed in two steps:

\begin{itemize}
\item We add unpolarized SI data (both SIDIS and hadronic data), with charge identification of $\pi^\pm$, $K^\pm$, and $p/\bar{p}$, to upgrade the NNFF1.0 FFs determined from SIA, and in particular separate out quark and anti-quark FFs. Once this is done, we could in turn use the same data to improve the NNPDF3.1 PDFs: here we expect changes to be small, except possibly in the strange sector, where $K^\pm$ SI processes could significantly improve the determination of $s$ and $\bar{s}$. Iteration to consistency should then be very rapid.

\item taking the FFs from the first step, we can now use polarized SI data, again with charge identification of $\pi^\pm$, $K^\pm$, and $p/\bar{p}$, to upgrade the NNPDFpol1.1 pPDFs, improving the separation of quark and anti-quark, and again with significant improvements expected in the polarized strange quark distributions. Again, once this is done, the polarised SI data can be used to improve the FFs, though here the impact should be relatively small. Consequently iteration to consistency should be even more rapid.

\end{itemize}

To carry out this sequence of fits, we would need to prepare FK tables for both unpolarized and polarized 
SI predictions, with either the (p)PDF or FF fixed to the appropriate central values, thus $\tilde{t}[f_0,d]]$, $\tilde{t}[f,d_0]$, $\tilde{t}[\Delta f,d_0]$, $\tilde{t}[\Delta f_0,d]$, with the corresponding covariance matrices $\tilde{s}^d[f_0]$, $\tilde{s}^f[d_0]$. However once this is done, the actual fitting should be no more challenging than in any other parton fit.

\section{Experimental prospects at the EIC}

\begin{figure}[t!]
\label{fig:eic-impact1}
\centering
\includegraphics[width=80mm]{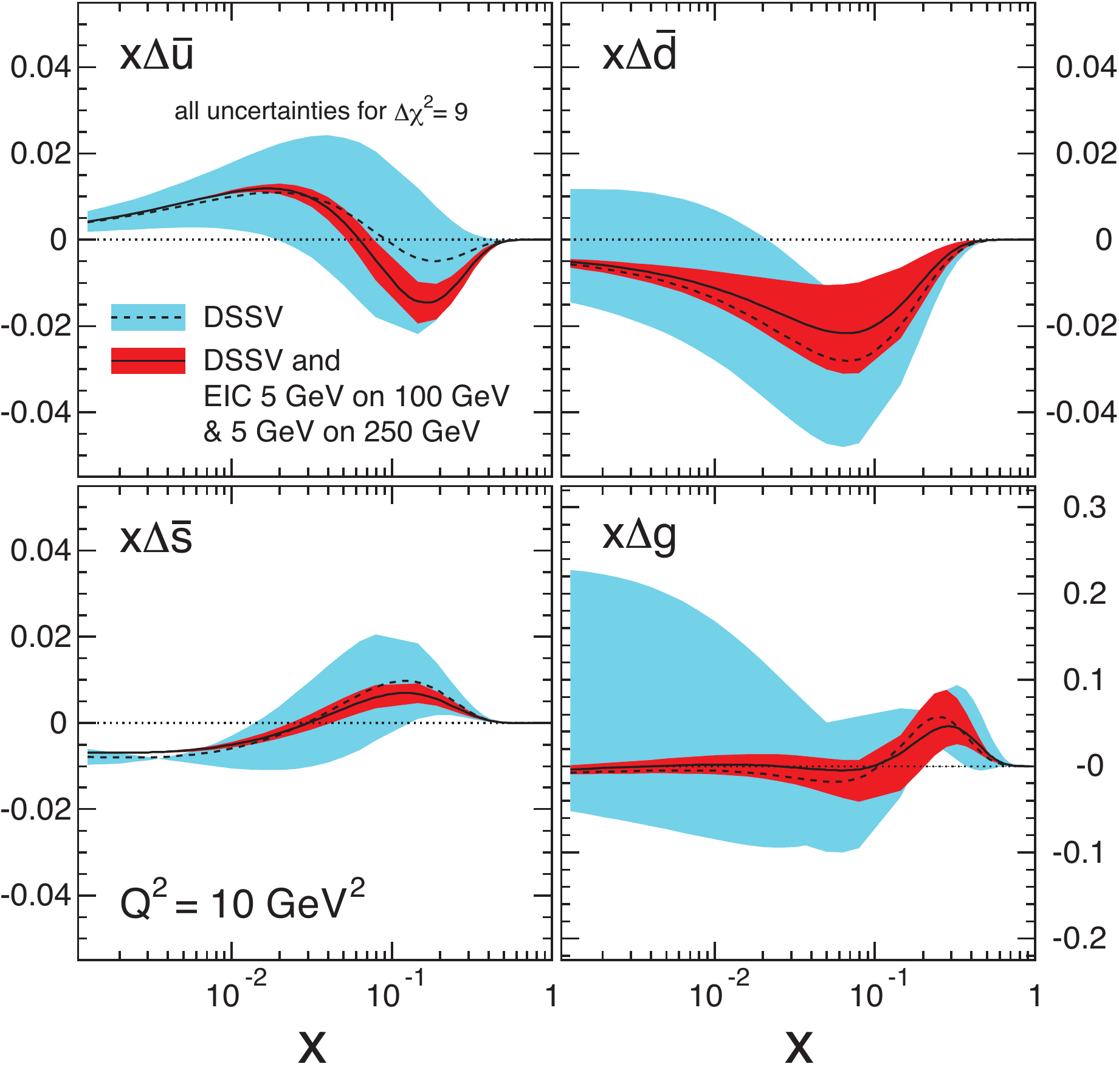}
\caption{The Uncertainty bands on polarized parton distributions in \MSb\ scheme, in the DSSV analysis (blue bands) and with EIC pseudo-data (red bands) using projected DIS and SIDIS EIC data expected in the future.}
\end{figure}

Due to its high luminosity and extended kinematic reach (see Fig.~\ref{fig:xq2}), the EIC~\cite{Deshpande:2005wd,Accardi:2012qut,Aschenauer:2014cki} offers the possibility of finally measuring the first moments of the parton helicities with sufficient accuracy to finally resolve the proton spin puzzle. Various studies have been 
performed~\cite{Aschenauer:2012ve,Aschenauer:2013iia,Ball:2013tyh} by adding simulations of EIC data to DSSV and NNPDFpol fits of pPDFs. Figure~\ref{fig:eic-impact1}
shows the results from the DSSV studies for the sea quark and gluon helicity distributions.  For comparison, the present uncertainty bands are also displayed. As one can see, an impressive
reduction in the width of the bands is expected from EIC data,
in particular, towards lower values of $x$.
Evidently, extractions of $\Delta g$ from scaling violations, and of
the light-flavor helicity distributions $\Delta u$, $\Delta d$ and
their anti-quark distributions from semi-inclusive scattering will be
possible with exquisite precision. With dedicated studies of kaon
production, the strange and anti-strange distributions would also be
accessible. 

\begin{figure}[t!]
\label{fig:eic-impact2}
\centering{\includegraphics[width=65mm]{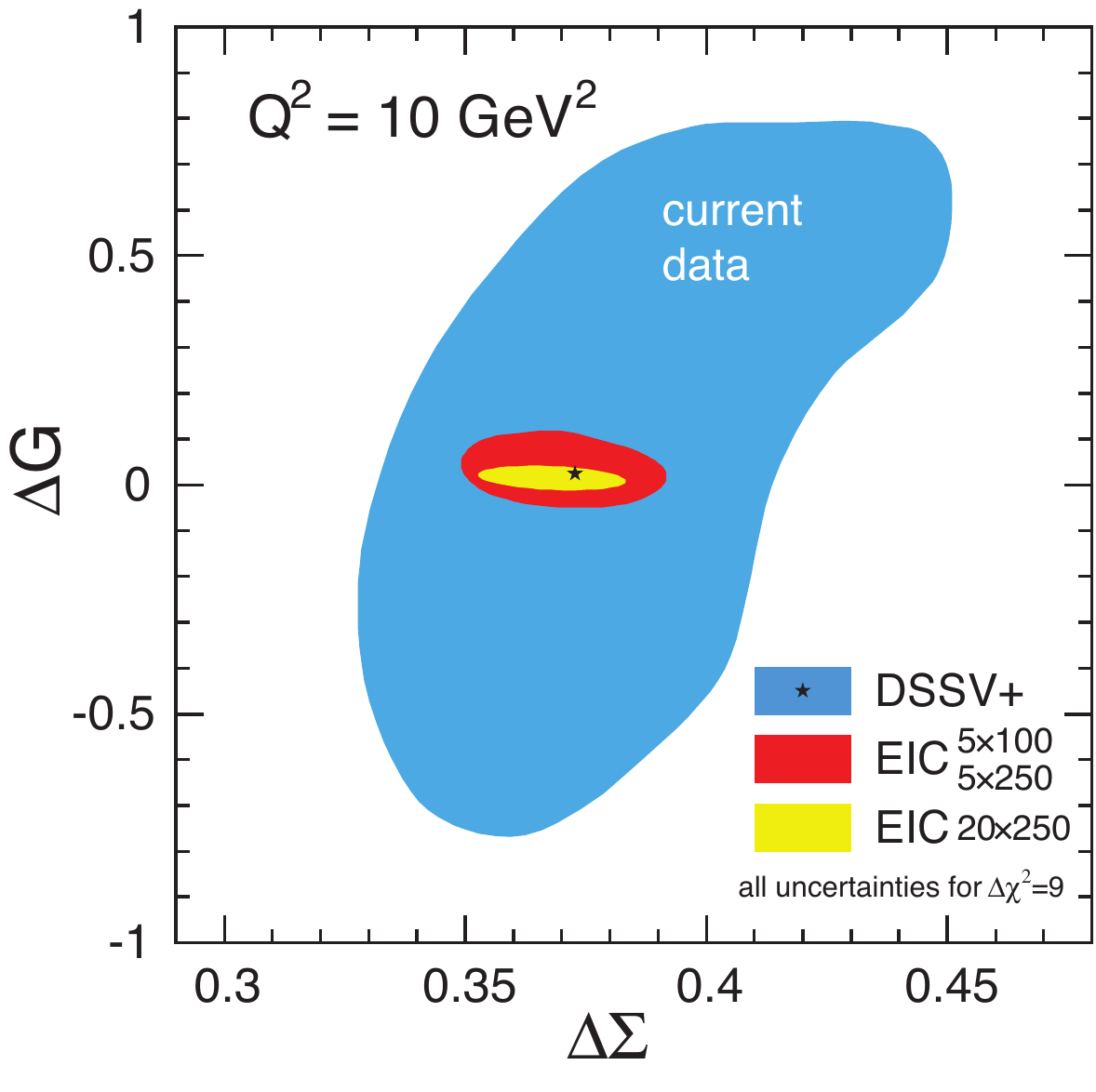}\vskip-0.5cm\hskip1.5cm
\includegraphics[width=105mm]{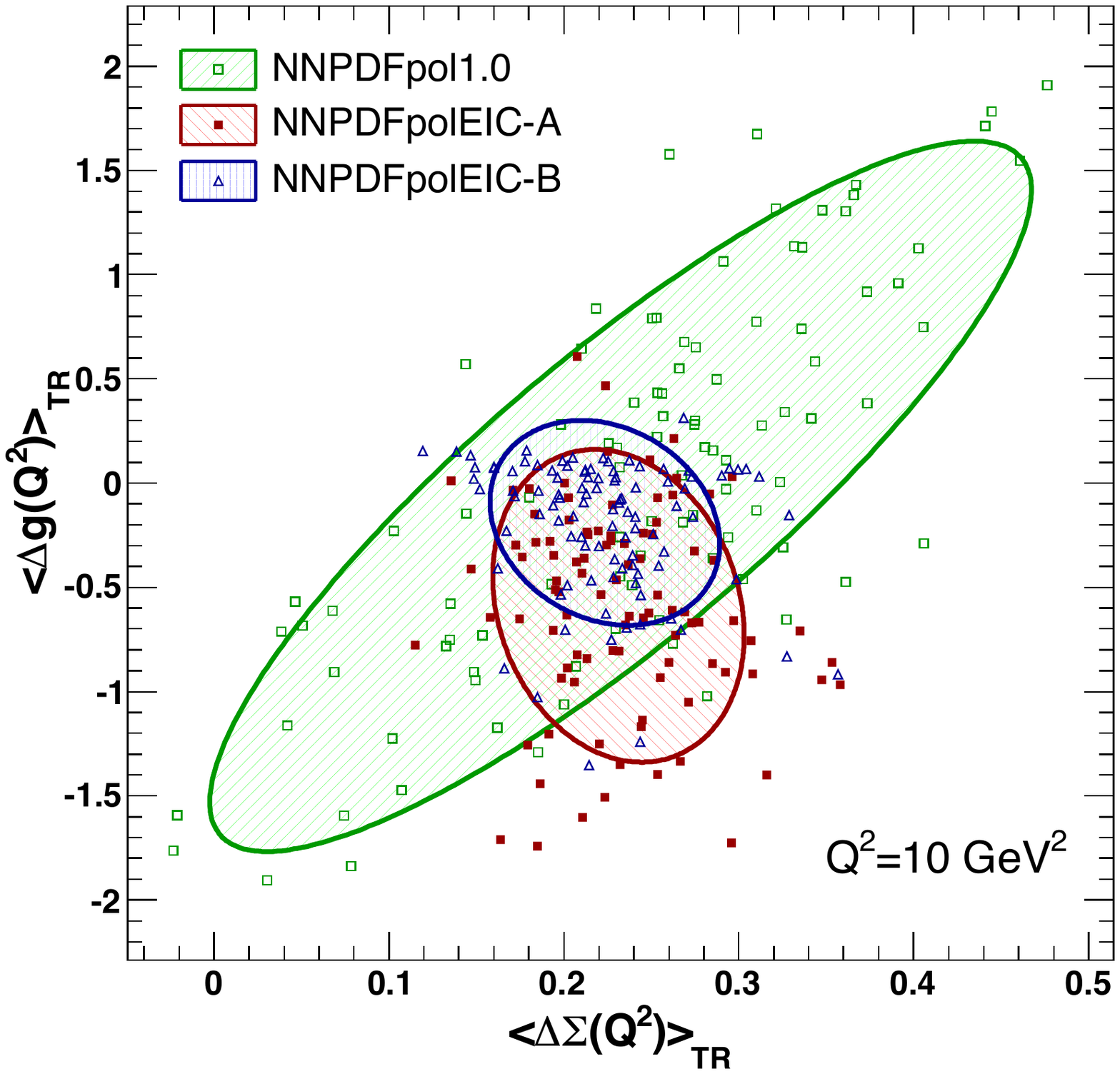}}
\vskip-0.5cm
\caption{The projected reduction in the uncertainties of the gluon helicity $\Delta g$ and the total the quark helicity $\Delta \Sigma$, comparing the current situation with what might be achieved by the EIC with operation at different beam energy combinations. The first moments are evaluated in the measured region $[0.001,1]$. The upper plot shows results from the DSSV analysis \cite{Aschenauer:2012ve}, current situation in blue, and projections in red and yellow. The uncertainties are determined with $\Delta \chi^2=9$. The lower plot shows similar results obtained by NNPDF \cite{Ball:2013tyh}, current situation in green, and projections in red and blue. The uncertainties are now genuine 1-$\sigma$ uncertainties obtained by averaging over replicas (also shown). Note the different scales on the two plots.}
\end{figure}

The projected reduction in the uncertainties of the gluon helicity $\Delta g$ and the total the quark helicity $\Delta \Sigma$ (in \MSb\ scheme, so this is actually $a_0(Q^2)$), comparing the current situation with what might be achieved by the EIC with operation at different beam energy combinations, is shown in Fig.~\ref{fig:eic-impact2}. Results are shown both for the DSSV analysis \cite{Aschenauer:2012ve} and from an independent analysis using NNPDF methodology \cite{Ball:2013tyh}. Although the results of the NNPDF analysis are more conservative, as might be expected, the general conclusions are the same: the anticipated kinematic range and precision of EIC data will give unprecedented insight into the size of the contributions $\Delta_g$ and $\Delta\Sigma$.  

Besides polarized proton beams, the EIC design envisions beams of
polarized deuterons or helium-3. 
The neutron's $g_1(x,Q^2)$ can thus
be determined, potentially with a precision that is comparable to the
data on $g_1(x,Q^2)$ of the proton.  The difference of the moments of
proton and neutron $g_1(x,Q^2)$ allows a test of the fundamental sum
rule by Bjorken~\cite{Bjorken:1968dy}.  The data from polarized fixed
target experiments have so far verified the sum rule to no better than
$20\%$ of its value \cite{Nocera:2014gqa}.  The extended kinematic range and improved
precision of EIC data will offer a more stringent tests of this sum rule,
as well as its corrections, to an accuracy that is expected to be driven mostly by advances in hadron beam polarimetry.

An additional, and unique, avenue for delineating the flavor structure
of the quark and anti-quark spin contribution to the proton spin at
the EIC is electroweak deep-inelastic scattering. At high $Q^2$, the
deep-inelastic process also proceeds significantly via exchange of $Z$
and $W^\pm$ bosons. This gives rise to novel structure functions that
are sensitive to different combinations of the proton's helicity
distributions. For instance, in the case of charged-current interactions
through $W^-$, the inclusive structure functions contribute, 
\begin{equation}\label{g5lo} g_1^{W^-}  = \Delta u +
\Delta\bar{d}+\Delta c +\Delta\bar{s} \; ,\qquad 
g_5^{W^-}= -\Delta u +
\Delta\bar{d}-\Delta c +\Delta\bar{s} \;,
\end{equation} 
\noindent
where $\Delta c$ denotes the proton's
polarized charm quark distribution.  The analysis of these structure
functions does not rely on knowledge of fragmentation.  Studies show
that both charged-current \cite{Aschenauer:2014cki} and neutral-current \cite{zhao:2017} interactions would be
observable at the EIC, even with relatively modest integrated
luminosities. To fully exploit the potential of the EIC for such
measurements, positron beams are required, albeit not necessarily
polarized. Besides the new insights into nucleon structure this would
provide, studies of spin-dependent electroweak scattering at short
distances with an EIC would be beautiful physics in itself, much in
the line of past and ongoing electroweak measurements at HERA,
Jefferson Laboratory, RHIC, and the LHC.  

\section{EIC Realization and Status}

The US nuclear scientific community meets approximately every five years to prepare a long range plan that is effective for ten subsequent years. In their last such planning exercise held in 2015, the community recommended construction of a high-energy, high-luminosity polarized electron-proton and electron-nucleus collider, now called the Electron Ion Collier (EIC) as the highest priority new construction facility. The primary aim of this machine that convinced the broader community to stand behind the recommendation included the role of gluons and sea quarks in building the nucleons and nuclei, and the partonic dynamics that underlies the interactions in QCD. Understanding the proton's spin structure, the principle aspect of the present article, is only one of the three major thrusts of the scientific goals of this future machine. To achieve the scientific program outlined in the science case the the machines designs should be able to deliver:
\begin{itemize}
\item Highly polarized ($\sim$ 70\%) electron and nucleon beams
\item Ion beams from deuteron to the heaviest nuclei (uranium or lead)
\item Variable center of mass energies from $\sim 40\, - \sim$100 GeV, upgradable to $\sim$150 GeV
\item High collision luminosity $\sim$10$^{33-34}$ cm$^{-2}$s$^{-1}$
\item Possibilities of having more than one interaction region
\end{itemize}

\begin{figure}[ht]
\label{fig:fig-eRHIC}
\centering
\includegraphics[width=90mm]{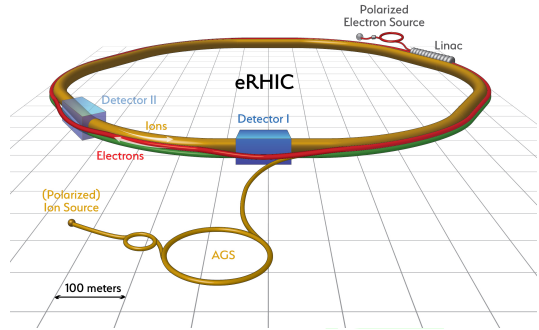}
\caption{The layout of eRHIC: The blue and yellow rings of RHIC exist, to which the electron beam facility would be added (shown in red) to realize electrion ion collisions in two experimental areas as shown.}
\end{figure}

\begin{figure}[ht]
\label{fig:fig-meic}
\centering
\includegraphics[width=75mm]{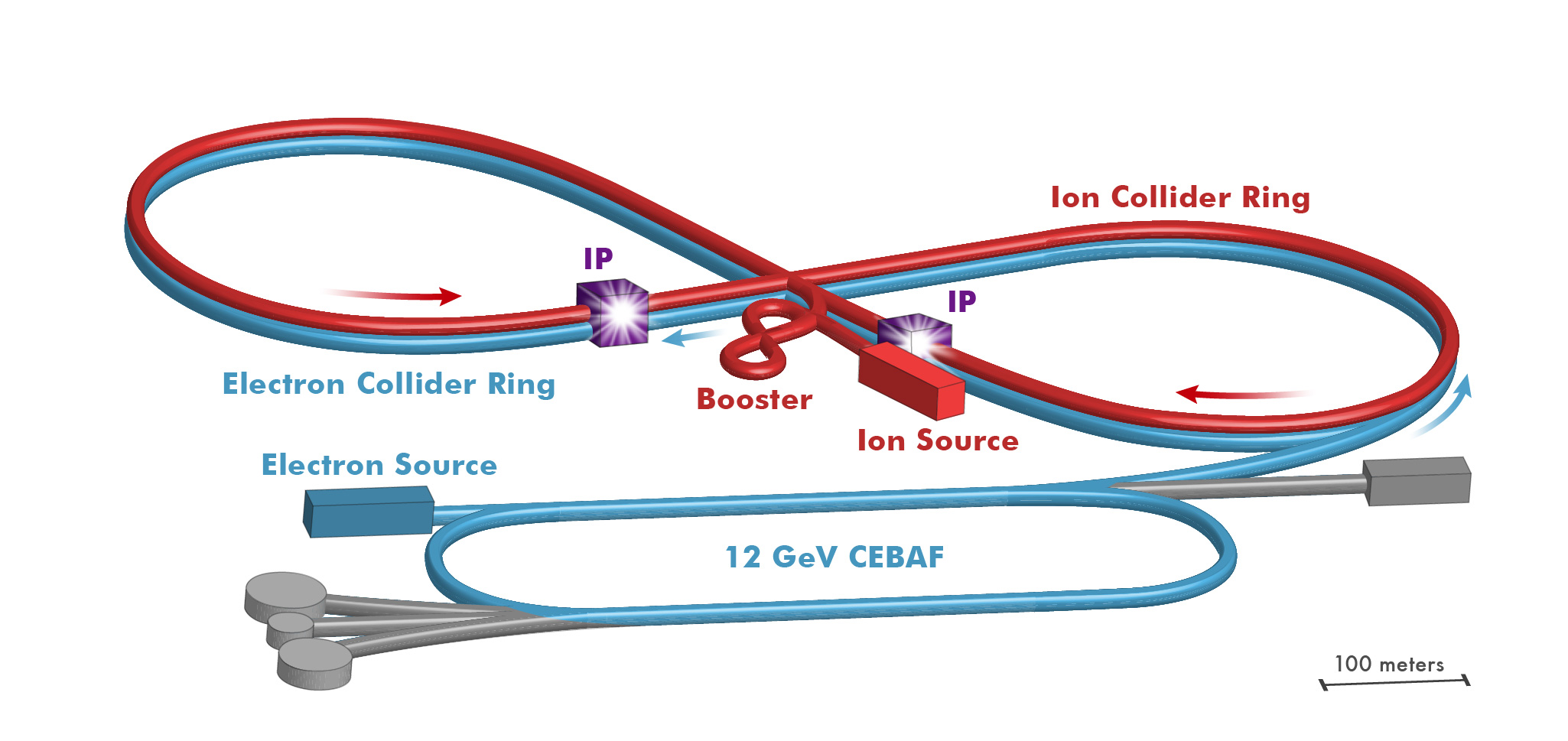}
\caption{The layout of JLEIC: The 12 GeV upgraded CEBAF in blue exists, to which the bow-tie (or 8) shaped complex with red (ion) and electron (blue) rings would be added, to realize collisions in two experimental areas as shown.}
\end{figure}

Currently there are two possible technical designs being considered by the DOE, based on existing infrastructure investments. One proposes to add an electron beam facility to the RHIC complex at BNL  to realize the EIC as eRHIC. A linac based design, ambitious in  its R\&D requirements has been worked out, and a more conventional ring-ring design is now being finalized as an option to reduce the uncertainties associated the challenges in the linac-ring design. The conceptual design of eRHIC is shown in Fig.~\ref{fig:fig-eRHIC}. 
The other uses the recently upgraded CEBAF at JLab as an electron beam injector in to a green field electron and hadron/nuclear beam complex to be built adjacent to CEBAF, in a novel bow-tie (or 8) shape and plans to achieve the above mentioned collisions parameters. The conceptual layout of JLEIC is shown in Fig.\ref{fig:fig-meic}. Both designs are expected to push the limits of our knowledge and control of accelerator parameters way beyond the state-of-the art, and hence are expected to be critical for future development and advancement in accelerator science. 

The physics-driven requirements on the EIC accelerator parameters and
extreme demands on the kinematic coverage for measurements makes
integration of the detector into the accelerator a particularly
challenging and yet absolutely essential feature of the EIC.  
Lessons learned from HERA have been considered while designing 
the EIC interaction region. Driven by the demand for high 
precision on particle detection and identification of final state particles in both $e$+$p$ and $e$+A programs, modern particle detector systems will be critical at the EIC. In order to keep the detector costs manageable, R\&D efforts are under way on various novel ideas for compact (fiber sampling and crystal) calorimetry, tracking (NaI coated GEMs, GEM size and geometries), particle identification (compact DIRC, dual radiator RICH and novel TPC) and high radiation tolerance for electronics. 

Motivated by the prospects of realization of EIC in the US, a world wide group of interested scientists have come together to form an EIC Users Group (eicug.org). Current participants not including students include $\sim$750 scientists, from $\sim$160 institutions and $\sim$29 countries around the world. It is expected that two interaction regions will provide for two independent detector systems built and operated by a world-wide user community forming at least two large experimental collaborations.

Considering the historical trends of US funding over the last 10 years, and assuming the same for the next 10 years, the Long Range Planning exercise suggests that the construction of the EIC could begin as early as 2022. Depending on which design is adopted (eRHIC or the JLEIC), one could then expect first collisions between 2027 and 2030. Currently the US National Academy of Science (NAS) is holding its own independent review of the EIC Science and is expected to give its verdict by Spring 2018. After a positive evaluation of its science the community expects a Critical Decisions Zero (CD0) status for the US EIC to be granted by the DOE. 

\section{Conclusions}

In this article we focused on one aspect of QCD and the possible physics at a future EIC, that of the helicity contributions from the quarks and gluons to the nucleon's spin. While there are now indications that gluons may carry a non-zero fraction of the proton's helicity, the situation is still inconclusive due to large uncertainties that come from unmeasured low-x regions. We have outlined two strategies for improving this situation in coming years: the systematic inclusion of polarized semi-inclusive data, to increase precision in the measured region, and improving the kinematic reach by polarized beam collisions at the future EIC. We still have a long way to go before we fully understand the partonic content and the dynamical origin of the proton spin in terms of the proton's constituents, however, we now know exactly what is needed to get us there.

\section{Acknowledgements}

RBD: I would like to take this opportunity to thank Guido, in absentia, for inspiration and support over many years. I first met Guido at Erice in 1983, and the impression he made then never left me. Many people's memories of Guido revolve around the slightly menacing question, calmly and simply expressed in his inimitable style, that identified precisely the weak point in any argument. However this was only one side of Guido: on a more personal level, he could also be both constructive and supportive. At the first scientific meeting of the Higgs Centre in Edinburgh in 2013, I was discussing with Guido some of the political difficulties we faced. His response was characteristically cryptic: ``the start is a most crucial step". Looking back now at his career, Guido made many crucial first steps: in QCD, Altarelli-Parisi, Altarelli-Ellis-Martinelli, Altarelli-Ross have each laid the foundations for an area of research that remains active to this day. It is generally much harder to start something than it is to end it, but at present there is no end in sight to the legacy of Guido Altarelli.

AD: I first met Guido in 1996 when he presented new data on $g_{1}^{d}$ from the SMC collaboration's NA47 experiment at the CERN theory seminar. Thereafter I remember fondly the discussions with Guido and Vernon Hughes about the need for a polarized electron-proton/deuteron collider. The idea of polarized HERA was very attractive then~\cite{Ball:1996de}, but unfortunately never happened due to the technical challenges of realizing polarized protons. However, after I mentioned the new idea of a polarized EIC (which used to be called eRHIC then) in the US, Guido strongly encouraged me to pursue it. Again his attitude was to encourage the crucial first step. Guido was very supportive of the RHIC spin program as a proof of principle for polarizing protons at high energy, making way for a future EIC. He would doubtless be very pleased that the EIC project, with high-energy polarized proton, light ion and electron beams has evolved to the point that we can be optimistic that it will actually get built. I would like to thank Guido for the enthusiasm that he showed for these new projects, and the support he gave to a young postdoctoral fellow like me.

RDB would also like to thank Luigi Del Debbio and Maria Ubiali for discussion of the implementation of theoretical uncertainties in PDF fits, and Emanuele Nocera for discussions about fragmentation functions.

\end{document}